%% file: main.tex
\documentclass{sigchi}

\usepackage{textcomp}

\newcommand{\system}{Evorus\xspace}
\newcommand{\eg}{{\em e.g.}\xspace}
\usepackage[usenames, dvipsnames]{xcolor}

\usepackage{amsmath}
\usepackage{amsmath}
\usepackage{booktabs}

\toappear{\scriptsize
Permission to make digital or hard copies of all or part of this work for personal or classroom use is granted without fee provided that copies are not made or distributed for profit or commercial advantage and that copies bear this notice and the full citation on the first page. Copyrights for components of this work owned by others than the author(s) must be honored. Abstracting with credit is permitted. To copy otherwise, or republish, to post on servers or to redistribute to lists, requires prior specific permission and/or a fee. Request permissions from \href{Permissions@acm.org}{Permissions@acm.org}.\\
\emph{CHI 2018}, April 21--26, 2018, Montreal, QC, Canada \\
\textcopyright~2018 Copyright is held by the owner/author(s). Publication rights licensed to ACM. \\
ACM 978-1-4503-5620-6/18/04...\$15.00  \\
\href{https://doi.org/10.1145/3173574.3173869}{https://doi.org/10.1145/3173574.3173869}

}

\usepackage{balance}       
\usepackage{graphics}      
\usepackage[T1]{fontenc}   
\usepackage{txfonts}
\usepackage{mathptmx}
\usepackage[pdflang={en-US},pdftex]{hyperref}
\usepackage{color}
\usepackage{booktabs}
\usepackage{textcomp}

\usepackage{microtype}        
\usepackage{ccicons}          

\usepackage{todonotes}

\usepackage{enumitem}

\newenvironment{choruschat}
    {
        \begin{enumerate}[leftmargin=3pc,style=nextline,align=right,label={}]
        \small
        \sffamily
    }
    {
        \end{enumerate}
    }

\newcommand{\crowd}[1]{\item [\textbf{crowd}]\xspace #1} 
\newcommand{\user}[1]{\item [\textbf{user}]\xspace #1} 
\newcommand{\voiceover}[1]{\item []\xspace [\textit{#1}]} 
\newcommand{\auto}[1]{\item [\textcolor{Violet}{\textbf{bot}}]\xspace \textcolor{Violet}{\textbf{#1}}} 

\def\plaintitle{\system: A Crowd-powered Conversational Assistant\\Built to Automate Itself Over Time}

\def\emptyauthor{}
\def\plainkeywords{crowd-powered system; crowdsourcing; real-time crowdsourcing; conversational assistant; chatbot}

\makeatletter
\def\url@leostyle{%
  \@ifundefined{selectfont}{
    \def\UrlFont{\sf}
  }{
    \def\UrlFont{\small\bf\ttfamily}
  }}
\makeatother
\def\pprw{8.5in}
\def\pprh{11in}

\usepackage{balance}  
\usepackage{graphics} 
\usepackage[T1]{fontenc}
\usepackage{txfonts}
\usepackage{mathptmx}
\usepackage{color}
\usepackage{booktabs}
\usepackage{textcomp}
\usepackage{microtype} 
\usepackage{ccicons}  

\usepackage{todonotes}

\usepackage[utf8]{inputenc}

\usepackage{amsmath}
\usepackage{enumitem}
\usepackage{framed}

\usepackage{booktabs}
\usepackage{multirow}

\usepackage{balance}  
\usepackage{graphics} 
\usepackage{times}    

\usepackage{xspace}

\usepackage[T1]{fontenc}
\usepackage{textcomp}
\usepackage[scaled=.92]{helvet} 
\usepackage{graphicx} 
\usepackage{balance}  
\usepackage{booktabs} 
\usepackage{ccicons}  
\usepackage{ragged2e} 
\usepackage{dialogue}

\usepackage{pifont}
\usepackage{blindtext}
\usepackage{scrextend}
\addtokomafont{labelinglabel}{\sffamily\small}

\newcommand{\kenneth}[1]{{\small\textcolor{blue}{\bf [#1 --Ken]}}}

\usepackage{enumitem}
\usepackage{url}

\usepackage{breakurl}

\usepackage{xspace}
\usepackage{balance}       
\usepackage{graphics}      
\usepackage[T1]{fontenc}   
\usepackage{txfonts}
\usepackage{mathptmx}

\definecolor{linkColor}{RGB}{6,125,233}
\hypersetup{%
  pdftitle={\plaintitle},
  pdfauthor={\emptyauthor},
  pdfkeywords={\plainkeywords},
  pdfdisplaydoctitle=true, 
  bookmarksnumbered,
  pdfstartview={FitH},
  colorlinks,
  citecolor=black,
  filecolor=black,
  linkcolor=black,
  urlcolor=linkColor,
  breaklinks=true,
  hypertexnames=false
}

\begin{document}

\title{\plaintitle}


\numberofauthors{1}
\author{%
  \alignauthor{\vspace{-2pc}Ting-Hao (Kenneth) Huang, Joseph Chee Chang, and Jeffrey P. Bigham\\
    \affaddr{Language Technologies Institute and Human-Computer Interaction Institute}\\
    \affaddr{Carnegie Mellon University}\\
    \email{\{tinghaoh,~josephcc,~jbigham\}@cs.cmu.edu}}
}


\maketitle

\begin{abstract}
\input{abstract}
\end{abstract}

\category{H.5.m.}{Information Interfaces and Presentation
  (e.g. HCI)}{Miscellaneous}


\section{Introduction}
\input{intro}

\section{Related Work}
\input{related-work}

\section{Evorus' Conversational Assistant Framework}
\input{system}

\section{Evorus' Automation and Learning Framework}
\input{system-learning}

\section{Part I: Learning to Choose Chatbots Over Time}
\input{bot-selection-model}

\section{Part II: Reusing Prior Responses}
\input{reusing-old-conv}

\section{Part III: Automatic Voting}
\input{auto-voting}




\section{Deployment Study and Results}


\input{deploy-0}
\input{deploy-1}

\input{deploy-2}


\section{Discussion and Conclusion}
\input{discussion}


\input{conclusion_and_future}


\balance{}

\bibliographystyle{SIGCHI-Reference-Format}
\bibliography{sample}

\end{document}

%% file: abstract.tex
Crowd-powered conversational assistants have been shown to be more robust than automated systems, but do so at the cost of higher response latency and monetary costs.
A promising direction is to combine the two approaches for high quality, low latency, and low cost solutions.
In this paper, we introduce \system, a crowd-powered conversational assistant built to automate itself over time
by {\em (i)} allowing new chatbots to be easily integrated to automate more scenarios,
{\em (ii)} reusing prior crowd answers,
and {\em (iii)} learning to automatically approve response candidates.
Our 5-month-long deployment with 80 participants and 281 conversations shows that \system can automate itself without compromising conversation quality.
Crowd-AI architectures have long been proposed as a way to reduce cost and latency for crowd-powered systems; \system demonstrates how automation can be introduced successfully in a deployed system. Its architecture allows future researchers to make further innovation on the underlying automated components in the context of a deployed open domain dialog system.

%% file: intro.tex
\begin{figure}[t]
  \centering
  \includegraphics[width=0.99\columnwidth]{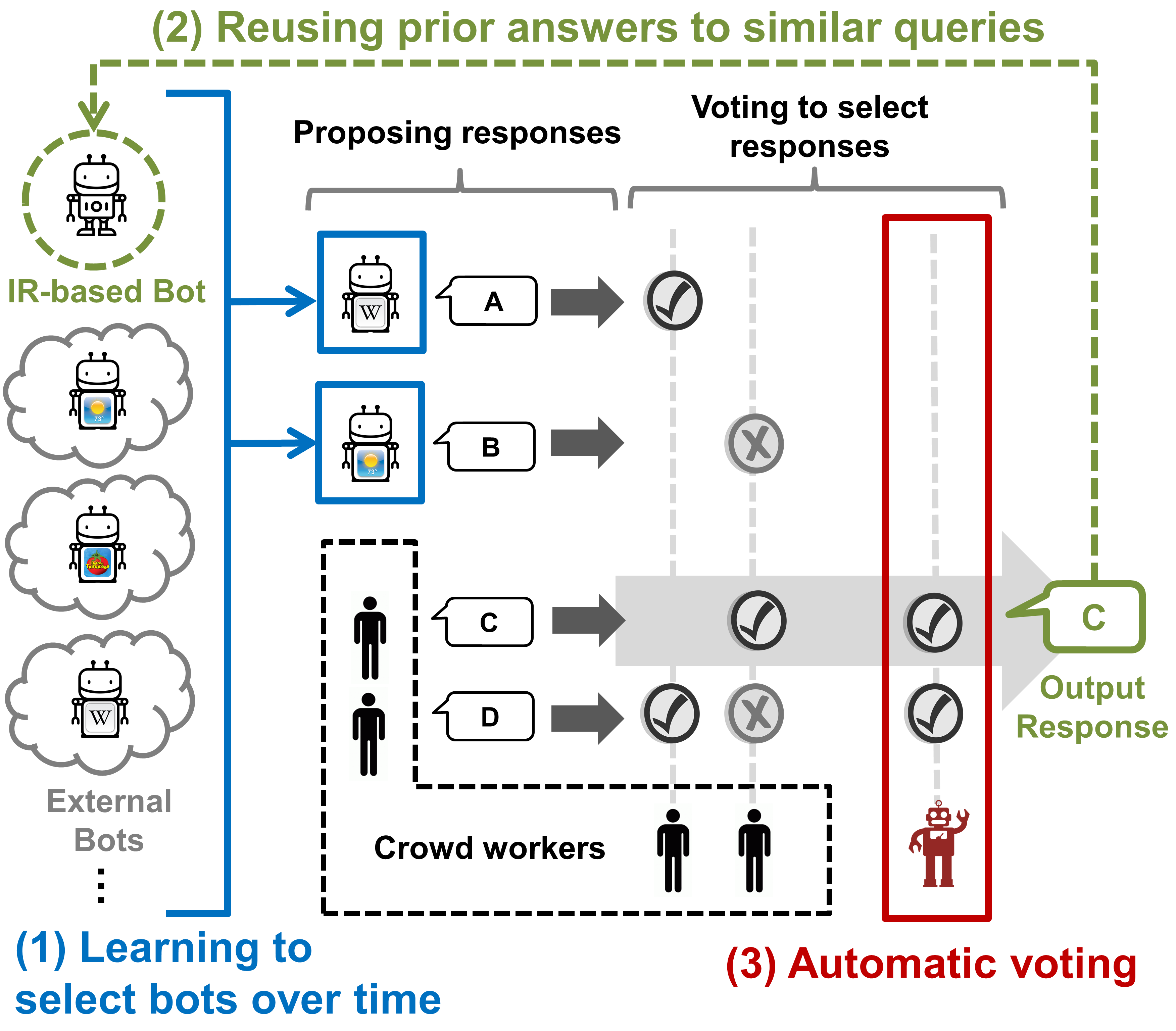}
  \vspace{-.3pc}
  \caption{\system is a crowd-powered conversational assistant that automates itself over time by {\em (i)} learning to include responses from chatterbots and task-oriented dialog systems over time, {\em (ii)} reusing past responses, and {\em (iii)} gradually reducing the crowd's role in choosing high-quality responses by partially automating voting.}
  \vspace{-1pc}
  \label{fig:system-overview}
\end{figure}

Conversational assistants, such as Apple's Siri, Amazon's Echo, and Microsoft's Cortana, are becoming increasingly popular, but are currently limited to specific speech commands that have been coded for pre-determined domains. As a result, substantial effort has been placed on teaching people how to talk to these assistants, {\em e.g.}, via books to teach Siri's language~\cite{talktosiri}, and frequent emails from Amazon advertising Alexa's new skills~\cite{meetAlexa}.
To address the problem of users not knowing what scenarios are supported, AI2 recently built an Alexa skill designed to help people find skills they could use,
only to have it rejected by Amazon~\cite{constine2017}.

Crowd-powered assistants are more robust to diverse domains, and are able to engage users in rich, multi-turn conversation. Some systems use professional employees,
such as Facebook M~\cite{facebookM},
while others use crowd workers, such as Chorus~\cite{chorus}. 
Despite their advantages, crowd-powered agents remain largely impractical for deployment at large scale because of their monetary cost and response latency~\cite{vizwiz,chorusDeploy}.
On the other hand, crowd-powered systems are often touted as a path to fully automated systems, but transitioning from the crowd to automation has been limited in practice. The most straightforward approach is to use data from prior conversations to train an automated replacement. This can work in specific domains~\cite{williams2007partially}, or on so-called ``chit-chat'' systems~\cite{banchs2012iris}. Fully automating a general conversational assistant this way can be difficult because of the wide range of domains to cover and the large amount of data needed within each to train automated replacements. Such automated systems only become useful once they can completely take over from the crowd-powered system. Such abrupt transition points mean substantial upfront costs must be paid for collecting training examples before any automation can be tested in an online system.

In this paper, we explore an alternative approach of a crowd-powered system architecture that supports gradual automation over time. In our approach, the crowd works with automated components as they continue to improve, and the architecture provides narrowly scoped points where automation can be introduced successfully. For instance, instead of waiting until an automated dialog system is able to respond completely on its own, one component that we developed recommends responders from a large set of possible responders that might be relevant based on the on-going conversation. Those responses are then among the options available to the crowd to choose. Another component learns to help select high-quality responses. Each problem is tightly scoped, and thus potentially easier for machine learning algorithms to automate.

This paper introduces \system, a crowd-powered conversational agent that provides a well-scoped path from crowd-powered robustness to automated speed and frugality.
Users can converse with \system in open domains, and the responses are chosen from suggestions offered by crowd workers and \textit{any number} of automated systems that have been added to \system. 
\system supports increased automation over time in three ways (Figure~\ref{fig:system-overview}): 
{\em (i)} allowing third-party developers to easily integrate automated chatterbots or task-oriented dialog systems to propose response candidates,
{\em (ii)} reusing crowd-generated responses from previous conversations as response candidates, and {\em (iii)} learning to automatically select high-quality response candidates to reduce crowd oversight over time.

In \system, existing dialog systems can be incorporated via simple REST (REpresentational State Transfer)
interfaces that take in the current
conversation context, and respond with a response candidate. Over time, \system
learns to select a subset of the automated components that are most likely to generate high-quality responses for different context. The responses are then forwarded to crowd workers as candidates. Workers then choose which of the responses to present to the users. \system sees 
workers selecting responses from candidates as signals
that enable it to learn to select both automated components and response candidates in the future.
It is important to note that while \system is a functioning and deployed system, we do not see the current version and its constituent components to be final. Rather, its architecture is designed to allow future
researchers to improve on its performance and the extent to which it
is automated, by working on constituent problems, which are each challenging in their own right. The structure of \system provides distinct learning points that can be bettered by other researchers. Others may include additional dialog systems or chatterbots, and improve upon its learning components, driven by the collected data and its modular architecture.

We deployed the current version of \system over time
to better understand how well it works.
During our deployment, automated response were chose 12.44\% of time, \system reduced the crowd voting by 13.81\%, and the cost of each non-user message reduced by 32.76\%.
In this paper, we explore when the system was best able to
automate itself, and present clear opportunities for future research to
improve on these areas.

\noindent This paper makes four primary contributions:




\begin{itemize}
\item {\bf \system Architecture:} a crowd-powered conversational assistant that is designed to gradually automate itself over time by including more responses from existent chatbots and reduce the oversight needed from the crowd;
\vspace{-.5pc}
\item {\bf Learning to Choose Chatbots Over Time:} we introduced a learning framework that uses crowd votes and prior accepted message to estimate the likelihood of each chatbots when receiving a user message;
\vspace{-.5pc}
\item {\bf Automatic Voting:} we implemented a machine learning model for automatically reducing the amount of crowd oversight needed, evaluated its performance on a dataset of real conversations, and developed a mathematical framework to estimate the expected reward of using the model; and
\vspace{-.5pc}
\item {\bf Deployment:} we deployed \system for over 5 months with 80 participants and 281 conversations to understand how the automatic components we developed could gradually take over from the crowd in a real setting.
\end{itemize}


%% file: related-work.tex
Our work draws from research on conversational agents, general purpose dialog systems and crowd-machine systems.

\textbf{General Purpose Dialog System:}
A number of general purpose dialog systems such as IRIS~\cite{banchs2012iris} have been proposed.
Wen et al.~\cite{wen2016multi} designed a neural network language generation model for multi-domain dialog systems.
A deep-learning-based domain adaptation model was also proposed recently~\cite{gasic2016dialogue}.
Project DialPort~\cite{zhao2016dialport} introduced a multi-agent framework that has the capability to include multiple task-oriented dialog systems to hold a multiple domain conversation.
On the other hand, in the field of natural language processing, general response generation technologies were also developed.
Ritter {\em et al.}~\cite{twitter-response} generated responses based on phrase-based statistical machine translation based on Twitter data.
Li {\em et al.}~\cite{li2016persona} introduced a response generator based on speaker model that encodes personas with background information and speaking style.
Recently, researchers started exploring end-to-end joint learning of language understanding in dialogue systems~\cite{chen2016end,yang2017end,li2017end}.
However, after decades of developments, sophisticated artificial ``conversational intelligence'' are largely absent in modern digital products.



\textbf{Crowd-powered Conversational Agents:}
Building fully-automated, open-domain conversational assistants is a widely researched topic in the field of artificial intelligence (AI), but has thus far remained an open challenge.
In response to this, the Chorus~\cite{chorus} conversational agent is powered by a crowd of human actors, which enables it to work robustly across domains~\cite{view}. 
To help users manage information and services through crowd-powered conversational agents, Guardian takes as input a Web API and a desired task from the user and the crowd determines the parameters necessary to complete the task~\cite{huang2015guardian,espGame};
IntructableCrowd helps users automate the management of sensors and tasks in their mobile phones through a conversational agent~\cite{Huang:2016:ICI:2851581.2892502}; and WearMail enables users to access their emails by talking to the crowd-powered assistant via smartwatch~\cite{wearmail2017}.
Conversational assistants powered by trained human operators, such as Facebook M~\cite{facebookM} and Magic Assistant~\cite{MagicAssistants}, have emerged in the recent years.  

While most crowd-powered conversational systems function well in laboratory settings, Chorus was deployed in the real world~\cite{chorusDeploy} and revealed a range of problems such as determining when to terminate a conversation or protecting workers from abusive content introduced by end users.
Microsoft Tey~\cite{tay} introduced an AI-powered agent which encountered problems when deployed publicly, because some users realized that they could influence what Tey would say because it mimicked them. Unlike Tey, \system does not learn only by mimicking users, and paid crowd workers are kept in the loop to verify responses in order to maintain quality. 
Our deployment did not reveal such problems and we believe the structure of \system makes such problems less likely.

\textbf{Crowd-Machine Hybrid Systems:}
Crowd and machine hybrid systems have enabled us to solve a wide range of tasks that were difficult for machines or humans to solve alone, making impacts in areas including crowdsourcing, machine learning, databases and computer vision \cite{Franklin:2011:CAQ:1989323.1989331,kamar2012combining,sarma2015surpassing,retelny2014expert,kamar2017complementing}. For instance, Flock and Alloy~\cite{cheng2015flock,chang2016alloy} use crowds to suggest predictive features, label data, and weigh these features with machine learning techniques to produce coherent categories and accurate models. The Knowledge Accelerator \cite{hahn2016knowledge} uses crowds to synthesize such crowd-machine structures into coherent articles. Zensors creates custom computer vision sensors bootstrapped by the crowd \cite{zensors}.
Similarly, CrowdDB~\cite{Franklin:2011:CAQ:1989323.1989331} uses human input for providing information that is missing from the database, for performing computationally difficult functions, and for matching, ranking, or aggregating results based on fuzzy criteria.
JellyBean~\cite{sarma2015surpassing} introduces a suite of crowd-vision hybrid counting algorithms that can perform in independent or hybrid modes returning more accurate counts that either workers or computer vision could do alone. 


%% file: system.tex

\begin{figure*}[t]
  \centering
  \includegraphics[width=0.91\textwidth]{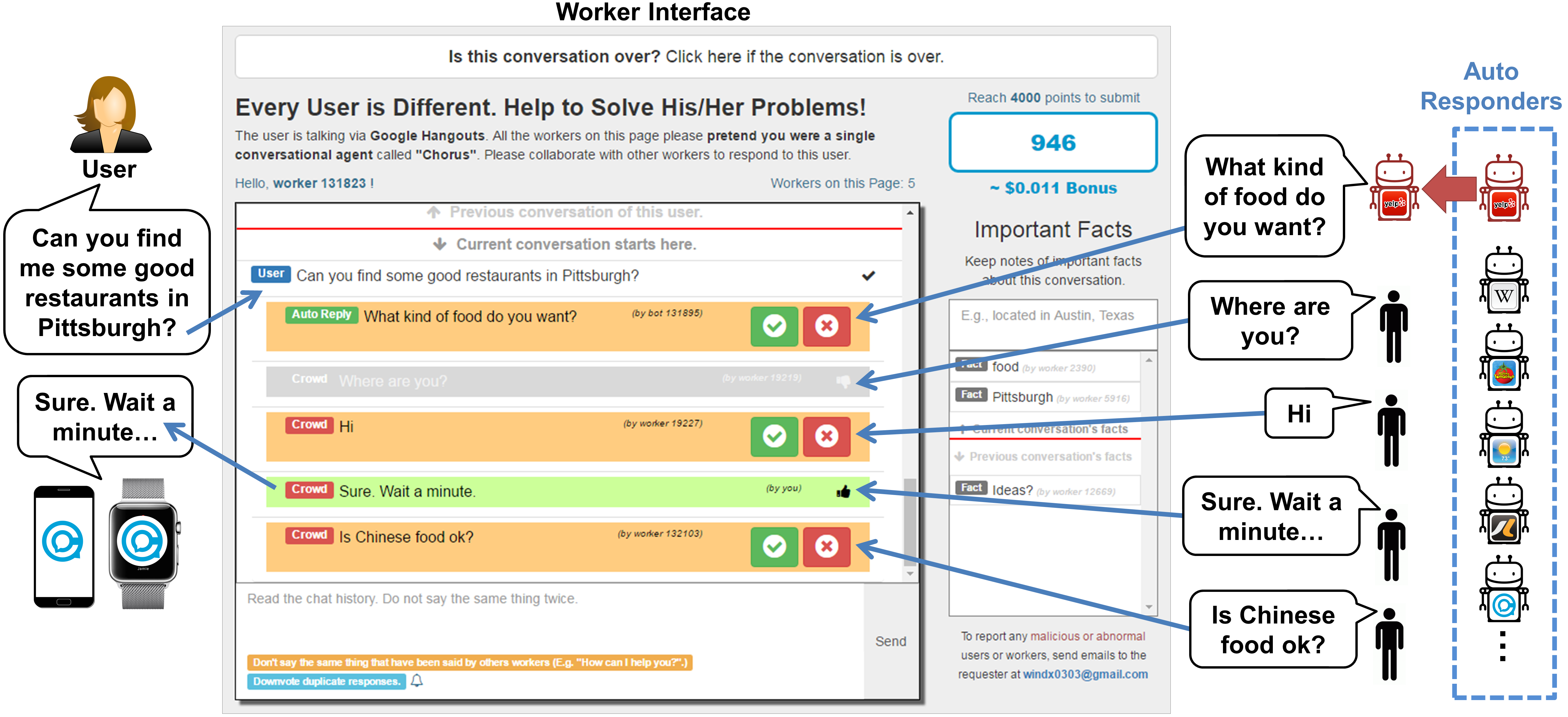}
  \caption{The \system worker interface allows workers to propose responses and up/down vote candidate responses. The up/down votes give \system labels to use to train its machine-learning system to automatically gauge the quality of responses. \system automatically expires response candidates upon acceptance of another in order to prevent workers from voting through candidate responses that are good but no longer relevant. Workers can tell each message is sent by the end-user (blue label), a worker (red label), or a chatbot (green label) by the colored labels.}
  \label{fig:ui-system}
  \vspace{-1pc}
\end{figure*}

\system obtains multiple responses from multiple sources, including crowd workers and chatbots, and uses a voting mechanism to decide which responses to send to the end-user.

\subsection{Worker Interface}

\system' worker interface contains two major parts (Figure~\ref{fig:ui-system}): the \textit{chat box} in the middle and the \textit{fact board} on the side.
Chat box's layout is similar to an online chat room.
Crowd workers can see the messages sent by the user and the responses candidates proposed by workers and bots.
The role label on each message indicates it was sent by the user (blue label,) a worker (red label,) or a bot (green label.)
Workers can click on the check mark (\ding{52}) to \textit{upvote} on the good responses, click on the cross mark (\ding{54}) to \textit{downvote} on the bad responses, or type text to propose their own responses.
Beside the chat box, workers can use the fact board to keep track of important information of the current conversation.
To provide context, chat logs and the recorded facts from previous conversations with the same user were also shown to workers.

The score board on the upper right corner displays the current reward points the worker have earned in this conversation.
If the conversation is over, the worker can click the long button on the top of the interface to leave and submit this task.

\subsection{Selecting Responses using Upvotes and Downvotes}


Crowd workers and bots can upvote or downvote on a response candidate.
As shown in Figure~\ref{fig:ui-system}, on the interface, the upvoted responses turned to light green, and the downvoted responses turned to gray. Crowd workers automatically upvote their own candidates whenever they propose new responses.
Upon calculating the voting results, we assigned a negative weight to a downvote while an upvote have a positive weights.
We empirically set the upvote weight at 1 and downvote's weight at 0.5, which encourages the system to send more responses to the user.
We inherited the already-working voting threshold from deployed Chorus~\cite{chorusDeploy}, which accepts a response candidate when it accumulates a vote weight that is larger or equal to 0.4 times number of active workers in this conversation.
Namely, \system accepts a response candidate and sends it to the user when Equation~\eqref{eq:thresh} holds:
\begin{equation}
\begin{gathered}
\#upvote \times W_{upvote} - \#donwvote \times W_{downvote} \\
\geqslant\#active\_workers \times threshold \\
W_{upvote} = 1.0, \quad W_{downvote} = 0.5, \quad threshold= 0.4\\
\end{gathered}
\label{eq:thresh}
\end{equation}
We formally defined the \textit{\#active\_workers} in the later subsection of real-time recruiting.
\system does not reject a message, so it does not have a threshold for negative vote weight.



\subsection{Expiring Unselected Messages to Refresh Context}

When \system accepts a response, the system turns the accepted message to a white background, and also \textit{expires} all other response candidates that have not been accepted by removing them from the chat box in the worker interface.
This feature ensures all response candidates displayed on the interface were proposed based on the latest context.
We also created a ``proposed chat history'' box on the left side of worker interface, which automatically records the worker's latest five responses.
Workers can copy his/her previously-proposed response and send it again if the message expired too fast.

\subsection{A Proposed, Accepted, or Expired Message}

In \system, non-user messages are in one of three states: [Proposed], [Accepted], or [Expired].
[Proposed] messages are open to be up/downvoted.
These messages were proposed by either a worker or a bot, has not yet received sufficient votes to be accepted, and has not yet expired;
[Accepted] messages received sufficient votes before they expired and were sent to the user;
and [Expired] messages did not receive sufficient votes before they expired. 
These messages were not sent to the users, and were removed from the worker interface.
A [Rejected] state does not exist since \system does not reject a message proactively.

\subsection{Worker's Reward Point System}

To incentivize workers, \system grants reward points to workers for their individual actions such as upvoting on a message or proposing a response candidate, and also for their collective decisions such as agreeing on accepting a message or proposing a message that were accepted.
The score box on the right top corner of the interface shows the current reward points to the worker in real-time.
Reward points are later converted to bonus pay for workers.
Without compromising output quality, if some of these crowd actions can be successfully replaced by automated algorithms, the cost of each conversation can be reduced.
\system' reward point schema was extended from the Chorus reward schema, which was previously used during its year-long deployment~\cite{chorusDeploy}.
This schema encodes the importance of each action, and thus provides a good guide for algorithms to estimate the benefit and risk when automating a crowd action.
Moreover, this reward schema will be used to estimate the expected reward points (and corresponding costs) an automatic voting bot can save, which we describe later.

\subsection{Real-time Recruiting \& Connecting to Google Hangouts}


When a conversation starts, \system uses the Ignition model \cite{ignition2017} to recruit workers quickly and economically from Amazon Mechanical Turk, and uses the Hangoutsbot~\cite{hangoutsbot_2017} library to connect with the Google Hangout servers so that users can use its clients to talk with \system on computers or mobile devices.
Each conversation starts with 1 worker and incorporates 5 workers at most.
Workers may reach a conversation at different times, but typically stay to the end of the conversation (average duration $\simeq$ 10 minutes).
The \textit{\#active\_workers} in Equation~\ref{eq:thresh} is defined as ``the number of crowd workers who were working on this conversation when the message was proposed,'' which varies as workers arrive (or drop out) at different times.
In our deployment,
the average \textit{\#active\_workers} of all crowd messages is 3.56 (SD=1.29,) and
77.56\% of the crowd messages had \textit{\#active\_workers} >= 3.

%% file: system-learning.tex
\system is a conversational assistant that is collaboratively run by real-time crowdsourcing and artificial intelligence.
The core concept of \system is to have crowd workers work with automated virtual agents (referred to as ``bots'') on the fly to hold sophisticated conversations with users.
To test this, we developed two types of bots: the automatic response generators, \emph{i.e.}, \textbf{chatbots}, and the automatic voting algorithms, {\em i.e.}, \textbf{vote bots}.
\system monitored all ongoing conversations, and periodically called chatbots and vote bots to participate in active conversations.
Both chatbots and vote bots take the entire chat log as input, and based on the chat log to generate responses or votes.
To coordinate with human workers' speed, \system often needs to set constraints on the frequency or capability (\eg, in which condition can a chatbot propose responses) of bots.
More importantly, \system can learn from the crowd feedback to automate itself over time via three primary mechanisms: {\em (i)} the chatbot selector, {\em (ii)} the retrieval-based chatbot that can reuse old responses, and {\em (iii)} the automatic voting bot.

\subsection{Part I: Learning to Choose Chatbots Over Time}

In \system, existing dialog systems or chatbots can be incorporated by defining a simple REST interface on top of them that accepts information about the current conversation state, and responses with a suggested response.
When a sufficient amount of chatbots are included in the bot pool, selecting the most appropriate chatbots to answer different questions becomes critical.
For instance, a simple ``ping-pong'' chatbot that always responds with what it was told can be selected to reply echo questions such as ``Hi'' or ``How are you?'';
a restaurant recommendation bot can be selected when the user is looking for food;
and a chatbot that was built on a friend's chat log can be selected when the user feels lonely~\cite{friendbot}.
In this paper, we introduce a learning framework that uses crowd votes and prior accepted messages to estimate the likelihood of each chatbot when receiving a user message.
The learning framework naturally assigns a slightly higher likelihood to newly-added chatbots to collect more data.
The beauty of this design is that any chatbot can contribute, as long as it can effectively respond to -- even a small -- set of user messages.

\subsection{Part II: Reusing Prior Answers}

Upon receiving a message from the user, \system uses a retrieval-based approach to find the most similar message in prior conversations and populates its prior response for the crowd to choose from.
By doing so, \system is capable to reuse the answer of prior similar questions to respond to users.
The advantage of using a retrieval-based method is that it naturally increases its capability of answering questions with the growth of the collected conversations, without the need of recreation or retraining of machine-learning models.
With the oversight of the crowd, the retrieval-based approaches also do not need to be perfect to contribute.
As long as it find good responses to a portion of user conversations, the learning framework described in Part I can gradually learn when to use it.

\subsection{Part III: Automatic Voting}

Closing the loop of automating the entire system, the last piece is to automate the oversight of the crowd that are necessary for quality control, {\em i.e.}, the voting process in \system.
We formulated response voting as a classification task and tackled it with a supervised machine-learning approach.
A set of features based on literature, including the word, the speaker, and the time of the proposed messages are used to develop a machine-learning model, and the prior collected crowd votes are used as gold-standard labels.
While the overall classifier performance is efficient in the dataset, a misfired vote (a false-positive) that mistakenly accepts a low-quality response will not only disturb the conversation, but also waste extra bonus money to crowd workers who proposed and voted for it.
In this paper we propose a mathematical framework to estimate expected benefits of using an automatic voting classifier.

In \system, both workers and the vote bot can upvote suggested responses. When a new suggestion is offered, the vote bot is called.
It first calculates its \textit{confidence} score, and, if the confidence is greater than a threshold, which is estimated by our proposed mathematical framework, the vote bot automatically upvotes the message.
\system monitors the latest down/upvotes and calculates voting results in real-time.
When a crowd message collects sufficient vote weight, \system \emph{(i)} accepts it and sends it to the user, and \emph{(ii)} removes all other candidate messages from the worker interface to refresh context.

In the following three sections, we describe in detail the three main parts of the \system framework.

%% file: bot-selection-model.tex



\system' chatbot selector learns over time from the crowd's feedback to choose the right chatbots to respond to user messages.
\system also \textbf{regularly populates lower-ranking chatbots} to allow the model to learn about new chatbots and tp keep the model up-to-date.

\subsection{Ranking and Sampling Chatbots}
Upon receiving a message from a user, \system uses both the text and prior collected data to estimate how likely each chatbot is capable of responding the user ({\em i.e.,} $P(bot|message)$).
We used a conditional probability, as shown in Equation~\ref{eq:bayesian}, to characterize the likelihood of selecting a chatbot ($bot$) after receiving a user $message$.
\begin{equation}
\label{eq:bayesian}
\begin{gathered}
P(bot|message) = P(bot) \times P(message|bot)\\
\approx P(bot) \times similarity(message,~history_{bot})
\end{gathered}
\end{equation}
$P(bot)$ is the prior probability of the chatbot, and $P(message|bot)$ is the likelihood of the user message given the chatbot's history (\emph{i.e.}, previous user messages that the bot has successfully responded).
While training an n-gram language model using previous messages to estimate $P(message|bot)$ is intuitive~\cite{Han:2011:LNS:2002472.2002520}, sufficient data for building such model is often unavailable for newly-added bots.
To generalize, we used a similarity measure based on distance between word vectors ($similarity(message,~history_{bot})$) to approximate this likelihood.
We will explain how we calculate these two components in the following subsection.

Equipped with the estimates, \system ranks all the chatbots based on the likelihood values, and always calls the first-ranking chatbot to provide its response.
More interestingly, in addition to the top chatbot, \system also randomly selects a lower-ranking chatbot to provide responses.
By doing so, \system is capable to gradually update its estimates of each bot based on the crowd feedback and learn over time the best scenario to call each chatbot.
Similar strategies, such as the \textit{epsilon-greedy strategy} that yields a small portion of probability for random outcomes and collects feedback,
have been used in models that learn to select crowd workers~\cite{tran2014efficient} and dialogue actions~\cite{Scheffler:2002:ALD:1289189.1289246}.
For the new chatbot, \system initially assigns a starting probability to it to allow the system to collect data about it, which we describe in the following subsection.

\subsection{Estimating Likelihood of a Chatbot}

We aimed at designing a learning framework that is {\em (i)} inexpensive to update, since we want the model to be updated every single time when the system receives a new label, and {\em (ii)} allows new bots to be added easily.

\textbf{Prior Probability of Chatbots:}
To generate more reliable prior estimation for newly-added bots with limited histories, we used a beta distribution with two shape parameters $\alpha$ and $\beta$ (Equation~\ref{eq:bot-prior}) to model the prior probability of each chatbot.
\begin{equation}
\label{eq:bot-prior}
P(bot) \approx \frac{(\text{\#accepted messages from bot})+\alpha}{(\text{\#user messages since bot online})+\alpha+\beta}
\end{equation}
$P(bot)$ can be interpreted as the \textit{overall acceptance rate} of the chatbot without conversation contexts.
The two shaping parameters $\alpha$ and $\beta$ can be viewed as the number of accepted (positive) and not-accepted (negative) messages that will be assigned to each new chatbot to begin with, respectively.
Namely, any new chatbot's prior probability $P(bot)$ will be initially assigned as $\alpha/(\alpha+\beta)$, and then later be updated over time.
The beta distribution's $\alpha$ and $\beta$ are both functions of the mean ($\mu$) and variance ($\sigma^2$) of the distribution.
In our pilot study, in which each automatic response requires only one vote to be accepted, four chatterbots had an average message acceptance rate of 0.407 (SD=0.028.)
Since we increased the required vote count from 1 to 2 in the final deployment, a lower acceptance rate is expected.
We used $\mu=0.3$ and $\sigma=0.05$ to estimate the shape parameters, where $\alpha=24.9$ and $\beta=58.1$.




\textbf{Similarity between Messages and Chatbots:}
To estimate $similarity(message,~bot)$, we first used the pre-trained 200-dimension GloVe word vector representation trained on Wikipedia and Gigaword~\cite{pennington2014glove} to calculate the average word vector of each message.
We then used previous user messages that were successfully responded by the chatbot as the bot vector $\vec{w}_{bot}$ that represents the chatbot in the word-vector space.
We also calculated the centroid vector of \textit{all} user messages, $\vec{w}_{overall}$, to represent general user messages.
Finally, as shown in Equation~\ref{eq:sim}, the similarity between a message and a chatbot is defined as the distance ratio between the vectors.
\begin{equation}
\label{eq:sim}
\begin{gathered}
similarity(message,~history_{bot})\\ \coloneqq
\frac{dist(\vec{w}_{message},\vec{w}_{overall})}{dist(\vec{w}_{message},\vec{w}_{bot})+dist(\vec{w}_{message},\vec{w}_{overall})}
\end{gathered}
\end{equation}
While $\vec{w}_{bot}$ can be  calculated as the centroid vector of prior user messages that were successfully responded to by the chatbot, in cold-start scenarios, a chatbot will not have sufficient accepted messages to calculate the vector.
We provide two solutions for chatbot developers:
First, the developer can provide a small set of \textbf{example messages} where their chatbots should be called.
For instance, the developer of an Yelp chatbot can list ``\emph{Find me a sushi restaurant in Seattle!}'' as an example.
\system will treat these example messages as the user messages that the chatbot successfully responded to, and use their centroid vector as the initial $\vec{w}_{bot}$.
When more messages are accepted, they will be added into this set and update the vector.
Second, for some chatbots, especially non-task chatterbots, it could be difficult to provide a set of examples.
Therefore, if the developer decided not to provide any example messages, we set the initial $dist(\vec{w}_{message},\vec{w}_{bot})=0$ for new chatbots.

%% file: reusing-old-conv.tex
\system uses an information-retrieval-based (IR-based) method to find answers to similar queries in prior conversations to suggest as responses to new queries.
To do so, \system first extracts query-response pairs from all the old conversations, and then performs a similarity-based sorting over these pairs. 

\textbf{Extracting Query-Response Pairs:}
One advantage \system has is that each accepted response has crowd votes, which can be used as a direct indicator of the response's quality.
For each turn of a conversation between one user and \system, we extracted the accepted crowd \textit{response} which did not receive any downvotes, along with the user message \textit{(query)} it responded to, as a \textit{(query, response)} pair.
Since the deployed \system has not had any prior conversation with users yet, we obtained conversation data that were collected by the deployed Chorus, which also used crowd voting to select responses, during May, 2016 to March, 2017 to start with.
We further removed the messages from known malicious workers and users, also removed all conversations where the users are co-authors or collaborators of Chorus~\cite{chorusDeploy}.
At the beginning of the \system deployment, 3,814 user messages were included, and each of these user message is paired with 1 to 5 crowd responses.

\textbf{Searching for the Most Similar Query:}
For the \textit{query} message in each \textit{query-response} pair, we calculated its average word vector by using the pre-trained 200-dimension GloVe word vector representation based on Wikipedia and Gigaword~\cite{pennington2014glove} and stores the vector in the database.
When \system receives a user message, the system first calculates its average word vector $\vec{w}_{message}$ using the same GloVe representation, and then searches in the database to look for the top $k$ responses that their corresponding queries' word vectors had the shortest Euclidean distances with $\vec{w}_{message}$.
Finally, for increasing answer's diversity, the system randomly selects one from top $k$ responses to send back to \system for the crowd to choose from.
We empirically set $k=2$ in our deployment.

%% file: auto-voting.tex
\system uses supervised learning to vote on responses.


\textbf{Data Preparation:} The voting mechanism has been proven to be useful in selecting good responses and holding conversations in the lab prototype~\cite{chorus} and deployed system~\cite{chorusDeploy}.
The final status ({\em i.e.}, accepted or not) of a messages is a strong signal to indicate its quality.
However, when we used this data to develop an AI-powered automated voting algorithm, it is noteworthy that expired messages were not all of lower quality.
In some cases the proposed response was good and fitted in the old context well, but the context changed shortly after the message was sent;
Some messages were automatically accepted and bypassed the voting process because \system does not have enough active workers, {\em i.e.}, when $0.4 \times \#active\_workers < 1$ in Equation~\ref{eq:thresh});
Furthermore, since downvotes can cancel out upvotes, the voting results in \system could be influenced by race conditions among workers.
For instance, when two upvotes of a message has been sent to the server, \system might decide that this message' vote weight is sufficient and sent it to the user, in which a belated downvote would deduct its vote weight to lower than the threshold.
Therefore, the training data needs to be carefully developed.

Similar to Part II,
we used voting data collected during the Chorus deployment~\cite{chorusDeploy} to train the initial machine learning model for voting.
We first extracted the expired messages with one or more downvote(s) as examples of ``downvote'', and extracted the accepted crowd messages with both one or more upvote(s) and zero downvote as examples of ``upvote.''
We excluded the automatically-accepted messages that were sent when the task did not have sufficient number of active workers ($0.4 \times \#active\_workers < 1$)
From all the messages collected by the deployed Chorus during September 2016 to March 2017, 1,682 ``upvote'' messages 674 ``downvote'' messages were extracted to from the dataset.

\textbf{Model \& Performance:} 
We then used this dataset to train a LibLinear~\cite{liblinear} classifier. 
For each message, \system extracted features in message, turn, and conversation levels to capture the dialogic characteristics~\cite{raux2008optimizing}, and also used GloVe word vector, which is identical as that of our retrieval-based response generation, to represent the content.
Our approach reached to a precision of 0.740 and a recall of 0.982 (F1-score = 0.844) on the ``upvote'' class in a 10-fold cross-validation experiment.
The feature analysis using the Weka toolkit~\cite{witten2016data} showed that the top 3 features were about the historical performance of the worker who proposed the message, and also 13 out of 20 top features were of a particular dimension of one of the word vectors.
On the other hand, the performance of the ``downvote'' class is less effective.
Its precision is 0.714 but recall is only 0.134.
This could be caused by the insufficient amount of training data, since Chorus promoted upvote more than downvote by design.
According to this result, in the deployed \system, the system only automatically upvoted when the classifier ouput ``upvote,'' but did not downvote otherwise.

\vspace{-.2pc}
\subsection{Optimizing Automatic Voting}
\begin{figure}[t]
  \centering
  \includegraphics[width=0.95\columnwidth]{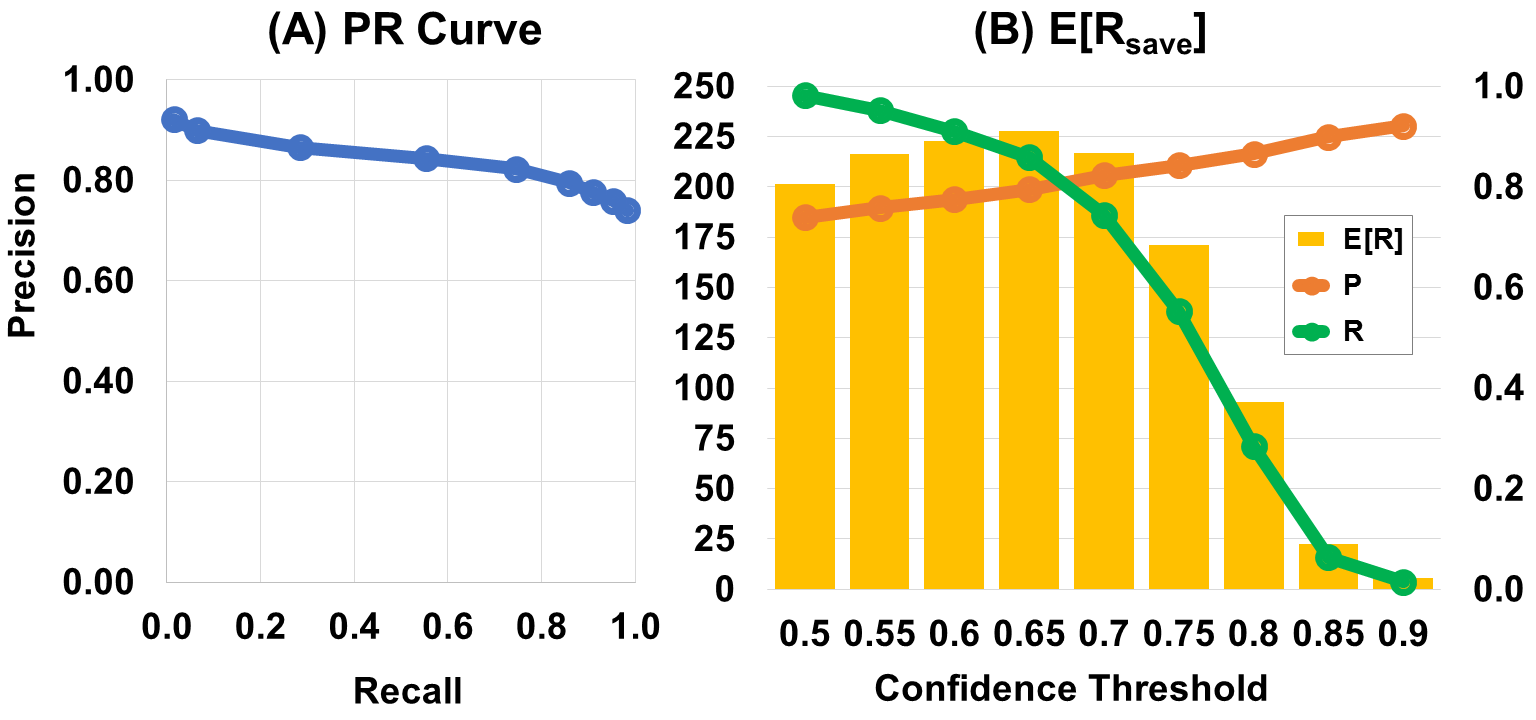}
 \vspace{-.3pc}
  \caption{(A) The precision-recall curve of the LibLinear classifier for automatic upvoting. (B) Using our model (Equation~\eqref{eq:general}) to estimate the precision and recall at different thresholds and their corresponding expected reward amount saved.}
  \label{fig:pr-curve}
  \vspace{-.5pc}
\end{figure}

Automatic voting directly participates in the process of deciding which messages to send.
While our machine-learning model resulted in good performance on our dataset,
we would like to use \system' worker reward point schema to find the right \textbf{confidence threshold} for the automatic voting classifier.
If the threshold is set too low, the classifier would vote frequently even when it is not confident about the prediction, and thus many low-quality responses would be accepted and disturb the conversation; if the threshold is set too high, the classifier would rarely vote, and the system will not gain much from using it.
Liblinear can output the probability estimates of each class when performing prediction, which we used as the notion of confidence of the classifier.
Thresholding out the predictions with lower confidences increased the precision but reduced the recall of the classifier.
Figure~\ref{fig:pr-curve}(A) shows the Precision-Recall curve of \system' upvoting classifier.

\textbf{Possible Outcomes When the Classifier Upvotes:}
To find confidence thresholds for \system, we introduced the following heuristics to estimate reward points saved per message by using the upvoting classifier.
Consider the following cases:


\begin{enumerate}
    \item \textbf{[Good Vote]} The classifier upvoted on a message that would originally be selected by the crowd. It saves 1 upvote reward ($R_{upvote} \times 1$) and 1 agreement reward ($R_{agreement} \times 1$) that would originally be granted to one human worker.
    \vspace{-.4pc}
    \item \textbf{[Bad Vote]} The classifier upvoted on a message that would originally \textit{not} be selected by the crowd. In this case, one of the two following consequences will occur: {\em (i)} \textbf{[Misfire]} The message is \textbf{sent} to the user. The system grants agreement rewards to all human workers who upvoted on this message ($R_{agreement} \times \#upvoted\_workers$) and 1 successful proposal reward to the worker who proposed the message ($R_{proposal} \times 1$); and {\em (ii)}: \textbf{[No Difference]} The message remains \textbf{not sent}. Even with one extra upvote, this message's vote count was still insufficient to get accepted. 
\end{enumerate}

\textbf{Estimating System's Expected Gain:}
Given these setups, the expected reward points $E[R_{save}]$ saved per message by using the classifier can be estimated as follows:
\begin{equation}
E[R_{save}]=TPR \times E[Good] - FPR \times E[Bad]
\label{eq:general}
\end{equation}
$TPR$ is the \textbf{true positive rate} and $FPR$ is the \textbf{false positive rate} of the classifier.
$E[Good]$ is the expected reward points saved per [Good Vote] event, and $E[Bad]$ is the expected reward points wasted per [Bad Vote] event.

In \system, $E[Good]$ is a constant ($R_{upvote} + R_{agreement}$).
Meanwhile, [Bad Vote] event costs reward points only when the upvoted message is sent ([Misfire]).
Therefore, $E[Bad]$ is decided by {\em (i)} how often one mistaken upvote triggers a misfire, and {\em (ii)} how expensive is one misfire, as follows:
\begin{equation*}
E[Bad] = P(\text{\emph{Misfire}}|Bad) \times E[R_\text{\emph{Misfire}}]
\end{equation*}
$P{(\text{\emph{Misfire}}|Bad)}$ is the conditional probability of a message being sent to the user given the classifier has mistakenly upvoted on it.
$E[R_{\text{\emph{Misfire}}}]$ is the expected reward points that were granted to workers in a single [Misfire] event. 

Our training dataset only used the not-accepted messages with at least one downvote to form the ``Downvote'' class, which were less likely to be misfired after adding one extra automatic vote.
For better estimating $P{(\text{\emph{Misfire}}|Bad)}$, we first ran the classifier on all messages that were \textit{not} included in our training set, and within all the messages that the classifier decided to upvote, we then calculated the proportion of messages that would be sent to the user if one extra upvote were added.
The rate was 0.692,
which we used to approximate $P{(\text{\emph{Misfire}}|Bad)}$.
Furthermore, based on \system' mechanism, $E[R_{\text{\emph{Misfire}}}]$ can be calculated as follows :
\begin{equation*}
E[R_{\text{\emph{Misfire}}}] = 
R_{agreement} \times E[\#upvoted\_workers] + R_{proposal}
\end{equation*}
$E[\#upvoted\_workers]$ is the expected number of human workers who upvoted on the message in an [Misfire] event.
Similarly,
we ran the classifier on the unlabelled data, and calculated the average number of workers who upvoted on the messages that the classifier decided to upvote on.
The number is 0.569 (SD = 0.731),
which we used to approximate $E[\#upvoted\_workers]$.


Finally, $E[Hit]=100+500=600$, and $E[R_{Bad}]=0.692 \times (500 \times 0.569 + 1000)=888.874$.
Using Equation~\eqref{eq:general}, we can estimate the precision and recall at different thresholds and their corresponding reward amount (Figure~\ref{fig:pr-curve}(B)).
According to the estimation, the best confidence threshold is at 0.65.
In the deployed \system we selected a slightly higher precision and set the threshold at 0.7 ($P=0.823$ and $R=0.745$.)

%% file: deploy-0.tex

\system was launched to the public as a Google Hangouts chatbot in March 2017.
While the end-users were not aware of the changes of the system from the client side, behind the scenes, our deployment had 3 phases: {\em (i)} Phase 1, {\em (ii)} Control Phase, and {\em (iii)} Phase 2.
Phase-1 deployment started in March, 2017.
We launched the system with only four chatterbots and one vote bot, without the learning component described in Part I, to understand the basics of having virtual bots working with human workers on the fly.
For comparison, in May 2017 we then temporarily turned off all automation components and had the system solely run by the crowd till late August 2017, which we referred to as the Control Phase.
Finally, for testing the capability of learning to select chatbots, we started Phase 2 deployment in early September.
The Phase-2 deployment included several significant changes: {\em (i)} increasing the frequency of calling chatbots for responses, {\em (ii)} increasing the vote count needed to accept a response from 1 to 2, and {\em (iii)} incorporating the Part I learning.

To recruit users, we periodically sent emails to mailing lists at several universities and posted on social media sites, such as Facebook and Twitter. 
Participants who volunteered to use our system were asked to sign a consent form first, and no compensation was offered.
After the participants submitted the consent form, a confirmation email was automatically sent to them to instruct them how to send messages to \system via Google Hangouts.
The users can use \system as many times as they want to, for anything, via any devices that are available to them.
Eighty users total talked with \system during 281 conversations.
The Phase-1 deployment had 34 users talked to \system during 113 conversations,
and Phase-2 deployment (till 17th September, 2017) had 26 users with 39 conversations.
The Control Phase had 42 users with 129 conversations.

%% file: deploy-1.tex



\subsection{Phase 1: Chatterbots \& Vote bot}

Our Phase-1 deployment explored how chatbots and our vote bot could work together synchronously with crowd workers. 
We implemented four chatterbots (including the IR-based chatterbot using Chorus conversation data described in Part II) and a vote bot.
During the Phase-1 deployment, the system only randomly selects one of four chatterbots to respond every half a minute, where the learning component described in Part I will be later included in Phase-2 deployment.



\textbf{Implementing Four Chatterbots:} In our Phase-1 deployment, we implemented the following four chatterbots.

\begin{enumerate}
  \setlength\itemsep{-0.2em}
  
    \item \textbf{Chorus Bot (shown as Part II):} A chatterbot that is powered by a retrieval-based method to reuse prior conversations to respond to users, which was described in Part II.
    \item \textbf{Filler Bot:} A chatterbot that randomly selects one response from a set of candidates, regardless of context.
    We manually selected 13 common ``conversation filler'' in the Chorus dataset (\eg, 
	``Is there anything else I can help you with?'', or
	``Thanks'') to form the candidate pool.
    \item \textbf{Interview Bot:} A chatterbot that uses a retrieval-based method, which is identical to Chorus Bot, to find the best response from 27,894 query-response pairs extracted from 767 transcripts of TV celebrity interview~\cite{cnn_trans}. 
        \item \textbf{Cleverbot:} Cleverbot is a third-party AI-powered chatbot which reuses more than 200 million conversations it had with users to generate responses~\cite{wiki:Cleverbot, cleverbot}. 
\end{enumerate}


\textbf{Vote Bot Setup:}
Currrently, 
The vote bot only votes on human-proposed messages, but not messages proposed by chatbots.
In Phase 1, \system requires each automatic vote to have at least one extra human upvote to be accepted.
Vote bots also skipped messages if the same worker proposed identical content earlier in the conversation but did not get accepted.
We believe that worker's re-sending is a strong signal of the poor quality of the message.
Vote bots can decide not to vote if the confidence is too low (Part III.)

\begin{figure}[t]
  \centering
  \includegraphics[width=0.95\columnwidth]{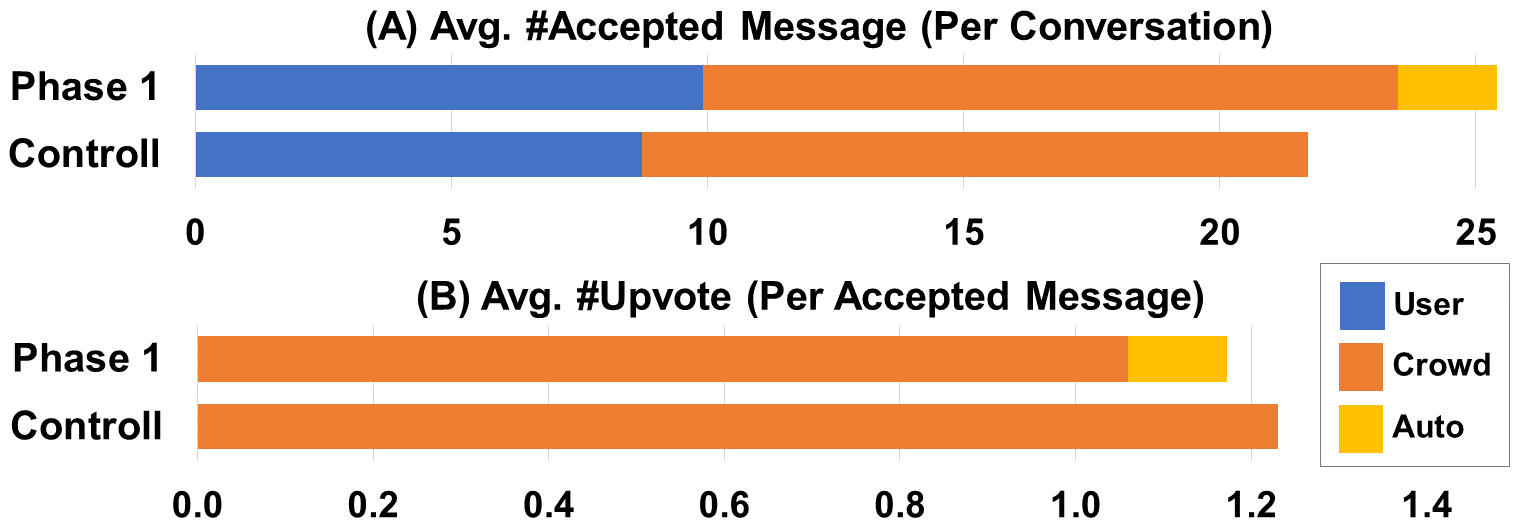}
  \vspace{-.3pc}
  \caption{Phase-1 Deployment: (A) Average number of accepted messages per conversation. Automated responses were chosen 12.44\% of the time. (B) Average number of upvotes per accepted non-user message. Human upvotes were reduced by 13.81\% by using automatic voting.}
  \label{fig:eval-compare}
\vspace{-1.1pc}
\end{figure}

\begin{figure*}[t]
  \centering
  \includegraphics[width=0.84\textwidth]{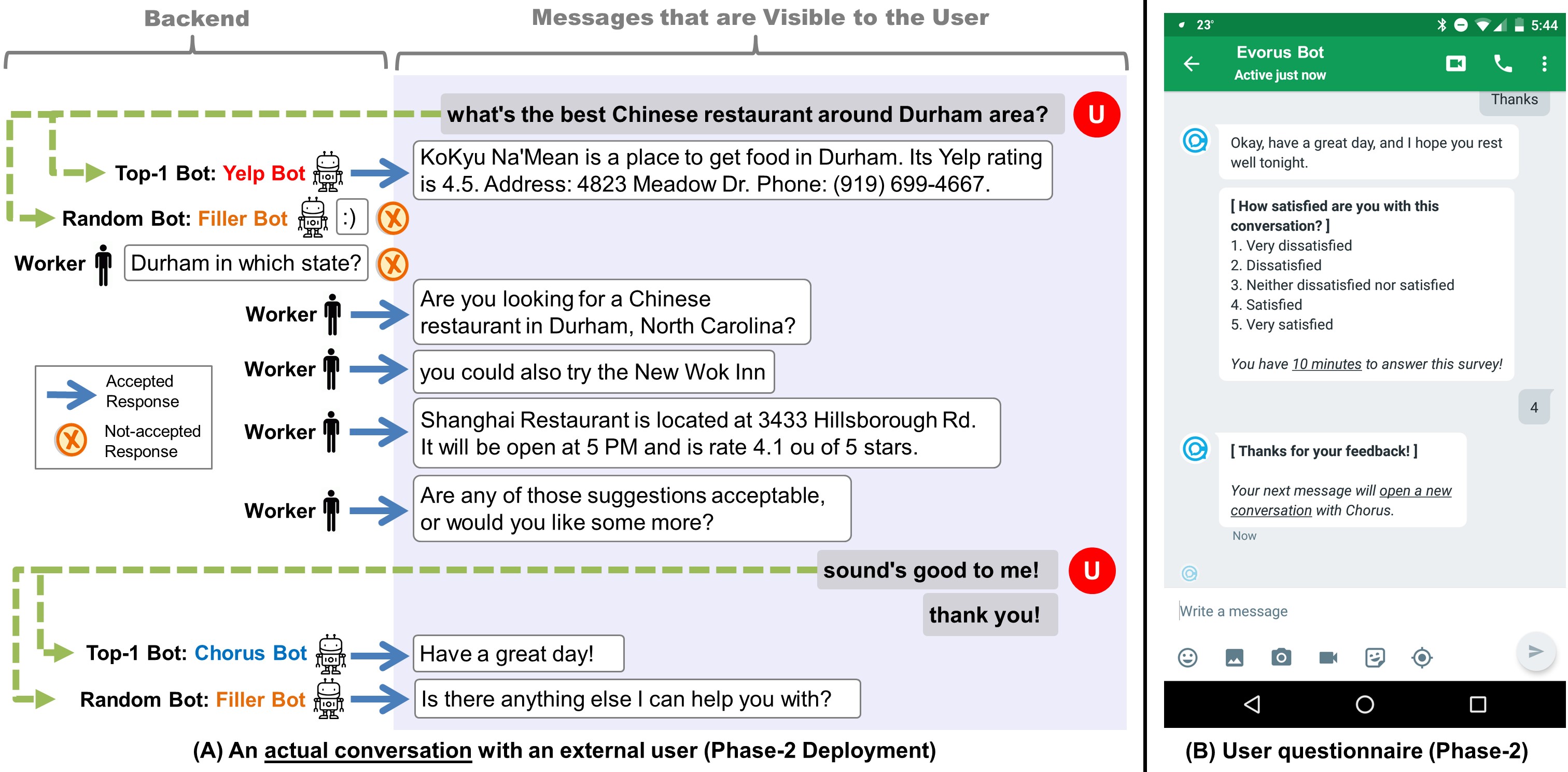}
  \vspace{-.3pc}
  \caption{(A) An actual conversation of \system. Conversations in \system tend to combine multiple chatbots and workers together. (B) User questionnaire used in Phase-2 deployment. The average user satisfaction rating of automated and non-automated conversations had no significant difference.}
  \label{fig:chatlog}
  \vspace{-.5pc}
\end{figure*}


\textbf{Automating Human Labors:}
During Phase-1 deployment, a conversation on average contained 
9.90 user messages (SD = 11.69), 
13.6 accepted messages proposed by the crowd workers (SD = 10.44), and 
13.58 accepted messages proposed by automatic chatterbots (SD = 2.81).
Thus, \textbf{automated responses were chosen 12.44\% of the time}.
As a comparison (Figure~\ref{fig:eval-compare}(A)), in the Control Phase (42 users and 129 conversations), a conversation on average contained 
8.73 user messages (SD = 10.05) and 12.98 accepted crowd messages (SD = 11.39).
In terms of upvotes, each accepted non-user message received 1.06 human upvotes (SD = 0.73) and 0.11 automatic upvotes (SD = 0.18)
In comparison, in the Control Phase,  
each accepted non-user message 
received 1.23 human upvotes (SD = 0.70.)
\textbf{Crowd voting was thus reduced by 13.81\%}.
The comparison is shown in Figure~\ref{fig:eval-compare}(B).
Moreover, an accepted non-user message sent by \system costed \$0.142 in Phase-1 deployment on average, while it costed \$0.211 during the Control Phase.
Namely, with automated chatbots and the vote bot, \textbf{the cost of each message is reduced by 32.76\%}.



We also calculated the acceptance rate of messages proposed by each chatbot.
The Filler Bot, which ignores context and proposes responses randomly, had the highest acceptance rate, 41.67\%.
The Chorus Bot's acceptance rate was 30.99\%, that of the Interview Bot was 33.33\%, and that of the Cleverbot was 30.99\%.
This might be because Filler Bot's commonplace responses ({\em e.g.}, ``I don't know'') were often considered acceptable by human raters.
In Phase 2, where an automatic response needed 2 human votes to go through, the Chorus Bot had the highest acceptance rate among all four chatterbots (Figure~\ref{fig:prior-change}).
While human workers, whose acceptance rate was 72.04\% during Phase 1, still outperformed all chatbots by a large margin, chatbots with low accuracy can still contribute to the conversation.
For instance, the Filler Bot, while being very simple and ignoring any context, nevertheless, often produces reasonable responses:


\begin{choruschat}
    \voiceover{The user asked information about the wildfire and smoke in Emory university campus.}
    \vspace{-.6pc}\user{Do you know where they are happening exactly? (The wildfires I mean)}
    \vspace{-.6pc}\auto{Can you provide some more details?}
\end{choruschat}


Compared with Filler Bot, Chorus Bot better targets its responses because it chooses messages based on similarity with previous human responses:


\begin{choruschat}
    \user{Hey~ how many people like bubble tea here?}
    \vspace{-.6pc}\auto{Ask for their feedback when you talk with them}
\end{choruschat}

\textbf{Conversation Quality:}
We sampled conversations with accepted automatic responses and a matching set without automated contributions.
For each, 8 MTurk workers rated [Satisfaction, Clarity, Responsiveness, Comfort], 
which was based on the PARADISE's objectives for evaluating dialogue performance~\cite{walker1997paradise} and the Quality of Communication Experience metric~\cite{liu2010quality}, on a 5-point Likert scale (5 is best.)
The original conversations (N=46) had an average rating of [3.47, 4.04, 3.88, 3.56], while those with automatic responses (N=54) had [3.57, 3.74, 3.52, 3.66].
The similar results suggest that the automatic components did not make conversations worse.

%% file: deploy-2.tex
\subsection{Phase 2: Learning to Select Chatbots}

Our Phase-2 deployment explored how the learning component, described in Part I, can select the right chatbots in context, and how integrating additional chatbots affects performance.
We implemented two additional \textit{utility bots}, a Yelp Bot and a Weather Bot, that can perform information inquiry tasks for different contexts, in addition to the four chatterbots and one vote bot from Phase 1.
We first launched the learning system with four chatterbots for two days, and then added the two utility bots for observing the changes of the model.
Furthermore, in order to directly compare user satisfaction levels, all the automated components (including the chatbots, the vote bot, and the learning component) were only applied to 50\% of the conversations randomly, and the other half of the conversations were solely run by the crowd as a baseline.

To efficiently collect crowd feedbacks to update our model, we increased the frequency \system called chatbots from randomly calling one chatbot and one vote bot every 30 seconds (Phase 1) to calling two chatbots (top-1 ranked plus random) and one vote bot every 10 seconds (Phase 2.)
To compensate for the possible drop in quality caused by higher calling frequency and randomly selecting one of the two bots, we increased the required upvote count for accepting an automatic response from 1 vote to 2 votes.
Namely, while \system obtained more automatic responses with a much higher frequency in Phase 2, it also required more human upvotes to approve each automatic response at the same time.
As a result, 
among the conversations that had automation in Phase 2, automated responses were chosen 13.25\% of the time, in which a conversation on average contained 
10.68 user messages (SD = 8.90), 
15.74 accepted messages proposed by the crowd workers (SD = 11.92), and 
2.40 accepted messages proposed by automatic chatterbots (SD = 2.40).
Each accepted non-user message received 1.90 human upvotes (SD = 1.13) and 0.30 automatic upvotes (SD = 0.22).
We did not compare Phase 2's results in detail to that of the Control Phase because the high-frequency setup of Phase 2 is primarily for experimental exploration.





	



\textbf{Implementing Two Utility Bots:}
In addition to the four chatterbots in the Phase-1 deployment, we implemented the following two task-oriented utility bots:
\begin{enumerate}
  \setlength\itemsep{-0.85em}
    \item \textbf{Yelp Bot:} A chatbot that suggests restaurants near the location mentioned by the user powered by the Yelp API~\cite{yelpApi}. If the user did not mention any location, it replies with ``You're looking for a restaurant. What city are you in?''.
    \item \textbf{Weather Bot:} A chatbot that reports the current weather of a mentioned city powered by the WeatherUnderground API~\cite{weatherApi}. If the user did not mention any city names, it replies, ``Which city's weather would you like to know?''
\end{enumerate}
The chatbots' developer (the first author) also provided three example chat messages for each that should trigger the corresponding chatbot, to initiate the learning process. For example, ``Any restaurant recommendations in NYC?'' for the Yelp Bot.

\textbf{Similar User Satisfaction Level:}
In Phase 2 we implemented an exit survey to measure end user satisfaction (Figure~\ref{fig:chatlog}(B)).
At the end of each conversation, the user had 10 minutes to report their satisfaction using a Likert-scale ranging from 1 (Very Dissatisfied) to 5 (Very Satisfied).
13 of 30 users provided feedback (response rate = 43\%).
The automated conversations' average user satisfaction rating was 4.50 (SD=0.5, N=4); the crowd conversations' average user satisfaction rating was 4.00 (SD=0.47, N=9), a difference that was not significant.

\begin{figure}[t]
  \centering
  \includegraphics[width=0.95\columnwidth]{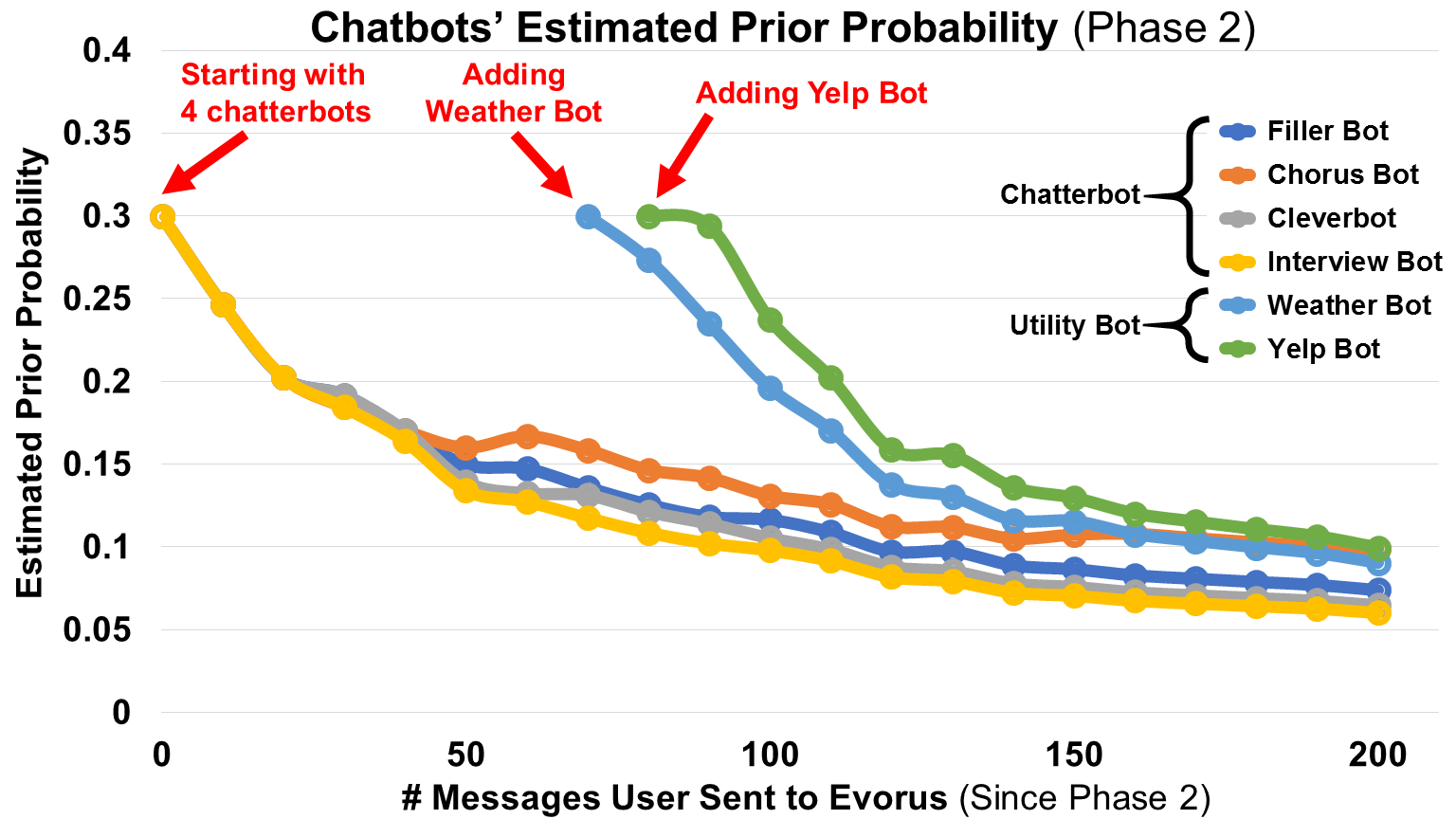}
  \vspace{-.2pc}
  \caption{The estimated prior probability (Equation~\ref{eq:bot-prior}) of each chatbot was continuously updated with the growth of user messages.}
  \label{fig:prior-change}
  \vspace{-.8pc}
\end{figure}

\textbf{Updating Estimates of Chatbot's Prior Over Time:}
While our deployment is of a medium scale, the dynamics of our likelihood model can still be observed.
For instance, the estimated prior probability described in Equation~\ref{eq:bot-prior} was continuously updated with the growth of conversation that \system had.
Our model assigned a starting probability of 0.3 to each chatbots (Figure~\ref{fig:prior-change}).
When users started talking with \system, crowd workers provided their feedback by upvoting and downvoting, and thus changed the estimation over time.
When new chatbots were added, \system intentionally assigned them a higher prior probabilities to allow quicker crowd feedback. 

\textbf{Utility Bots in Cold-Start Scenarios:}
We would like to understand if \system can select appropriate chatbots to obtain responses in corresponding context.
Since non-task chatterbots such as the Cleverbot could be difficult for humans to judge if it should be called given a message,
we focused only on task-oriented utility bots in the evaluation.
For each user message in the automated conversations in Phase 2, the researchers manually annotated if it is relevant to the topic of ``weather'' and ``restaurant,'' respectively.
With the assumption that each utility bot should be called when its topic comes up, we compared the human-labelled topic against the top chatbot that were suggested by \system and calculated the precision, recall, and F1-score. 
It is noteworthy that we only evaluated when the appropriate chatbot was called, regardless of the quality of the responses it generated.
As a result, two newly-added utility bots both had a high precision and a lower recall.
The Weather Bot's precision was 1.00 and the recall was 0.47 (F1=0.64);
and the Yelp Bot's precision was 0.67 and the recall was 0.20 (F1=0.31.)
This result shows the nature of new bots in our learning framework: \system uses the expert-generated small set of examples to calculate the initial vector of each chatbot, which could result in precise predictions but with lower coverage.
Over time, when \system collects more examples that each bot has successfully responded to and the recall increases.
Moreover, conversations in \system tend to combine multiple chatbots and workers together. For instance, in the following conversation, the crowd had the user to narrow down the query, and then the Weather Bot was able to answer it.
An additional detailed example is shown in Figure~\ref{fig:chatlog}(A).


\begin{choruschat}
\vspace{-.3pc}
    \voiceover{The user asked about the weather in Afghanistan.}
    \vspace{-.5pc}\crowd{What city in Afghanistan?)}
    \vspace{-.5pc}\voiceover{The crowd sends a website about Afghanistan weather.}
    \vspace{-.5pc}\user{Kabul}
    \vspace{-.5pc}\auto{Friday's weather forecast for [Kabul, Afghanistan]: Cloudy with a few showers. High 79F. Winds NW at 5 to 10 mph. Chance of rain 30\%.}
\vspace{-.3pc}
\end{choruschat}

\textbf{Error Cases:}
Our bot-selection algorithm starts with high precision and low recall, and increases recall as it gradually gathers examples.
Therefore, most errors we observed were false-negatives, where a chatbot should have been triggered but was not.
Other errors came from the chatbot, where the bot was correctly triggered but its response was invalid.
Workers usually downvoted or ignored these suggestions.
In the rare cases where invalid automatic responses were mistakenly sent to the user, the crowd often tried to explain how the response was automatically generated to the user afterward.

\begin{choruschat}
\vspace{-.3pc}
    \user{hi! Can you summarize the features of the new iPhone for me?}
    \vspace{-.5pc}\voiceover{Multiple messages list the features of the iPhone X.}
    \vspace{-.5pc}\auto{There is no iPhone 7.}
    \vspace{-.5pc}\voiceover{The crowd lists more features, and the user says thank you.}
    \vspace{-.5pc}\crowd{No problem!}
    \vspace{-.5pc}\crowd{Some Auto replies don't even make sense}
\vspace{-.3pc}
\end{choruschat}


			

%% file: discussion.tex

We introduced \system, a crowd-powered system conversational assistant built to automate itself over time. Informed by two phases of public field deployment and testing with real users, we iteratively designed and refined its flexible framework for open-domain dialog.
We imagine a future where thousands of online service providers can develop their own chatbots, not only to serve their own users in a task-specific context, but also to dynamically integrate their services into \system with the help of the crowd, allowing users to interact freely with thousands of online services via a universal portal.
Supporting this scale offers opportunities for future research.
For example, one direction is to improve the learning framework to support third-party chatbots that also improves overtime, or to better balance between the exploitation and exploration phases (like in a multi-armed bandit problem). \system could also be used to collect valuable fail cases to enable third-party developers to improve their bots ({\em i.e.}, when a bot was triggered, but its proposed response was rejected).

\system has three main advantages as compared to previous approaches.
First, it is a working system that can serve as a scaffold for automation over time.
A core advantage of starting with a working system is that users can talk to \system naturally from day one, ensuring conversation quality while collecting training data for automation.
Second, given the oversight of the crowd, \system has a high tolerance for errors from its automated components.
Even an imperfect automation component ({\em e.g.,} chatbots) can contribute to a conversation without hurting quality, which yields more space for algorithms to ``explore'' different actions ({\em e.g.,} selecting a chatbot with medium confidence.)
Finally, 
\system allows a mixed group of humans and bots to collaboratively hold open conversations.


Most automated systems created from crowd work simply use the crowd for data; \system tightly integrates crowds and machine learning, and provides specific points where automated components can be introduced. 
This architecture allows each component to be improved, providing a common research harness on which researchers specializing in different areas may innovate and compete.
For instance, ``response generation'' has long been developed in the NLP community; \system provides a natural evaluate it within a larger conversational system.
The flexibility of the \system framework potentially allows for low cost integration between many online service providers and fluid collaboration between chatbots and human workers to form a single user-facing identity.
Given the complexity of conversational assistance, \system is likely to be crowd-powered in part for some time, but we expect it to continue to increasingly rely on automation.

%% file: conclusion_and_future.tex
\section{Acknowledgements}
We thank Walter S. Lasecki and 
also the workers who participated in our studies. This project has been funded as part of the Yahoo!/Oath InMind Project at Carnegie Mellon University.